\documentclass[aps,prl,a4paper,notitlepage,reprint,superscriptaddress]{revtex4-1}

\usepackage{amsmath,amssymb,amsfonts} 
\usepackage{mathrsfs} 

\usepackage{mathtools} 
\usepackage{color}

\usepackage{graphicx,float}
\usepackage[colorlinks,
linkcolor=red,
citecolor=blue,
urlcolor=red]{hyperref}

\usepackage{cleveref}

\renewcommand{\t}[1]{\mathrm{#1}}

\newcommand{\figref}[1]{Fig.~\ref{#1}}
\renewcommand{\eqref}[1]{Eq.~\ref{#1}}

\begin{document}
\title{Fractal-like mechanical resonators with soft-clamped fundamental mode}
	
\author{S. A. Fedorov}
\email{sergey.fedorov@epfl.ch}
\affiliation{Institute of Physics (IPHYS), {\'E}cole Polytechnique F{\'e}d{\'e}rale de Lausanne, 1015 Lausanne, Switzerland}

\author{A. Beccari}
\affiliation{Institute of Physics (IPHYS), {\'E}cole Polytechnique F{\'e}d{\'e}rale de Lausanne, 1015 Lausanne, Switzerland}

\author{N. J. Engelsen}
\affiliation{Institute of Physics (IPHYS), {\'E}cole Polytechnique F{\'e}d{\'e}rale de Lausanne, 1015 Lausanne, Switzerland}
	
\author{T. J. Kippenberg}
\affiliation{Institute of Physics (IPHYS), {\'E}cole Polytechnique F{\'e}d{\'e}rale de Lausanne, 1015 Lausanne, Switzerland}
\email{tobias.kippenberg@epfl.ch}
	
\begin{abstract}
Self-similar structures occur naturally and have been employed to engineer exotic physical properties. Here we show that acoustic modes of a fractal-like system of tensioned strings can display increased mechanical quality factors due to the enhancement of dissipation dilution. We describe a realistic resonator design in which the quality factor of the fundamental mode is enhanced by as much as two orders of magnitude compared to a simple string with the same size and tension. Our findings can open new avenues in force sensing, cavity quantum optomechanics and experiments with suspended test masses. 
\end{abstract}
	
\date{\today}
\maketitle
	 

\paragraph{Introduction ---} 
Self-similar structures can have unusual physical properties. Coast lines are a famous example---their length is loosely defined at geographic scale\cite{mandelbrot_how_1967}. In the domain of optics, it was found that self-similar cavities can support modes with arbitrarily small mode volume\cite{choi_self-similar_2017} at a given wavelength. Meanwhile, hierarchical metamaterials can have improved stiffness per unit mass\cite{lakes_materials_1993,rayneau-kirkhope_ultralight_2012} compared to natural materials. The acoustic vibrations of resonators are also known to be affected by structural self-similarity in a nontrivial way, both in terms of the vibrational mode density\cite{alexander_density_1982,sapoval_vibrations_1991} and damping\cite{sapoval_acoustical_1997}. The latter can aid the design of mechanical resonators with low dissipation.

\begin{figure}[t]
\centering
\includegraphics[width=\columnwidth]{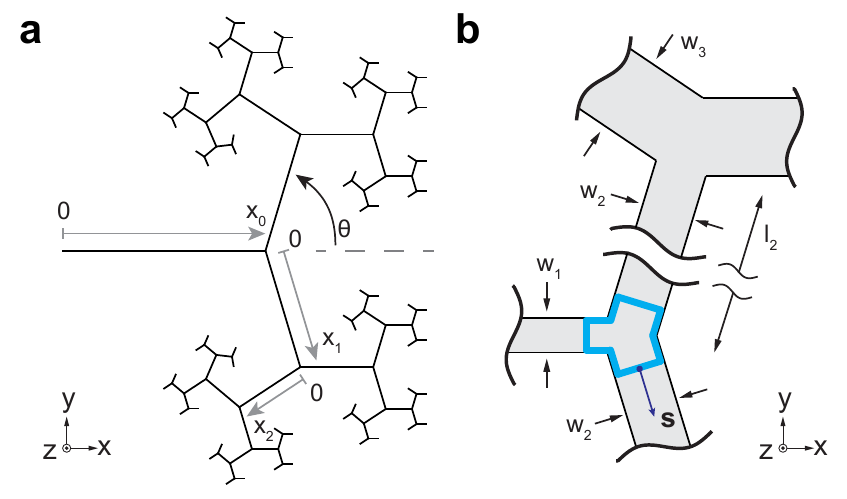}
\caption{\small{
a) Binary tree with six branching levels, local $x$-coordinates are shown for the first three levels. One tree defines half a resonator, the complete structure is formed by adding its mirror reflection in the $yz$ plane. b) Two branching points of a binary tree resonator with the definitions of the segment widths, $w$, and lengths, $l$. The blue contour is used to derive the transformation of the mode derivative.}}
\label{fig:intro}
\end{figure}

In this work we study mechanical vibrations of systems of tensioned strings in the shape of self-similar binary trees, which are clamped at the tips in order to sustain tension (see \figref{fig:intro}a). In such structures, due to the combination of high aspect ratio and static stress, the intrinsic loss-limited quality factors ($Q$s) of flexural modes are controlled by dissipation dilution\cite{gonzalez_brownian_1994,unterreithmeier_damping_2010,schmid_damping_2011}. The diluted $Q$ of a resonator mode is related to the material loss angle $\phi$ (the delay between stress and strain) as\cite{gonzalez_brownian_1994,unterreithmeier_damping_2010,schmid_damping_2011,yu_control_2012,fedorov_generalized_2019}
\begin{equation}\label{eq:Q}
Q= D_Q/\phi.
\end{equation}
The dissipation dilution coefficient, $D_Q$, can be much greater than one and is found as\cite{gonzalez_brownian_1994,unterreithmeier_damping_2010,schmid_damping_2011,yu_control_2012,fedorov_generalized_2019} 
\begin{equation}\label{eq:DQ}
D_Q=\frac{\langle W_\t{total}\rangle}{\langle W_\t{lossy}\rangle}=\frac{1}{\alpha \lambda +\beta \lambda^2}.
\end{equation}   
Here $\langle W_\t{total} \rangle$ is the dynamic elastic energy averaged over the vibrational period, $\langle W_\t{lossy}\rangle$ is its lossy part, and the parameter $\lambda=h/l\sqrt{E/(12\sigma)}$ depends on the resonator length, $l$, thickness in the direction of deformation, $h$, Young's modulus, $E$, and static stress, $\sigma$. The two terms in the denominator of \eqref{eq:DQ}, which scale differently with $\lambda$, come from the integration of lossy energy over the resonator's clamped boundary and the bulk. We call $\alpha$ and $\beta$ the boundary and distributed loss coefficients, respectively.

In a self-similar binary tree, flexural modes reduce in amplitude as they propagate from the trunk to the clamped tips of the branches. This can suppress the boundary loss coefficient, $\alpha$, which limits the dissipation dilution factor $D_Q$ in high aspect ratio structures. To date, the only known way of fully suppressing $\alpha$, or making a mode ``soft-clamped'' according to the terminology introduced in Ref.~\cite{tsaturyan_ultracoherent_2017}, is the localization of a mode by a phononic crystal (PnC) shield\cite{tsaturyan_ultracoherent_2017,ghadimi_strain_2017}. This technique, in combination with other recently developed methods for engineering dissipation dilution\cite{tsaturyan_ultracoherent_2017,ghadimi_strain_2017,fedorov_generalized_2019,bereyhi_clamp-tapering_2019,fedorov_generalized_2019,sadeghi_influence_2019}, resulted in the demonstration of quality factors up to $10^9$\cite{ghadimi_strain_2017,rossi_measurement-based_2018} in nanomechanical resonators made of stoichiometric Si$_3$N$_4$. However, localization by PnC can only be applied to high order vibrational modes---the aforementioned record $Q$s were demonstrated for mode orders in the range from tens to hundreds. Achieving similar $Q$ for the fundamental resonator mode, which is well separated from the rest of the spectrum, would be advantageous in many applications.

The manuscript is structured as follows. First we show how the propagation of a flexural mode of a string across a single branch point reduces the boundary loss coefficient. Next we present a theory of binary-tree resonators which use cascaded branchings to enhance quality factors. Using this theory we calculate the $Q$s of an example Si$_3$N$_4$ nanoresonator and predict, in excellent agreement with finite element simulation, a fundamental mode $Q$ exceeding $10^9$ assuming experimentally realistic parameters. We then study the variation of boundary and distributed loss coefficients with the parameters of the binary trees and show that the trade-off for a reduction of boundary loss is an increase in the distributed loss.  

Our results are relevant to areas ranging from sensing\cite{garcia-sanchez_casimir_2012,fischer_spin_2019} to cavity quantum optomechanics\cite{aspelmeyer_cavity_2014}, which employ stressed, high-$Q$ nanomechanical resonators\cite{verbridge_high_2006,zwickl_high_2008,unterreithmeier_universal_2009}. Moreover, because of the close relationship between the $Q$ of the fundamental mode of a clamped tensioned structure and the $Q$ of a pendulum\cite{gonzalez_brownian_1994}, our results can be used for designing high-$Q$ suspensions of test masses, akin to those employed in gravitational wave detectors and experiments on macroscopic optomechanics\cite{corbitt_measurement_2006,corbitt_optical_2007,bodiya_sub-hertz_2019}.

\paragraph{Boundary losses ---}  In order to show how boundary lossy elastic energy can be suppressed in a binary tree we review the relation between the boundary loss coefficient, $\alpha$, and the mode amplitude gradient. The transverse displacement field $u(x)$ of a flexural string mode in the high-tension limit is found from the equation
\begin{equation}\label{eq:waveBeam}
-\sigma(x)\frac{d^2 u}{dx^2}=\omega^2 \rho u,
\end{equation}   
where $x$ is the coordinate along the string, $\sigma(x)\equiv \sigma_{xx}(x)$ is the static axial stress distribution, $\rho$ is the material density and $\omega$ is the mode frequency. \eqref{eq:waveBeam} is valid everywhere except in the regions close to the string clamping points. Once the standing wave approaches a clamping point, it enters a transition region over which it reduces its gradient to zero. In the vicinity of the clamped boundary, the mode experiences sharp bending with curvature given by (see SI of Ref.~\cite{fedorov_generalized_2019}, also \cite{gonzalez_brownian_1994,yu_control_2012})
\begin{equation}\label{eq:ucl}
u''_\t{cl}(x)=\frac{u'(+0)}{\lambda_\t{cl}l}\exp\left(-\frac{x}{\lambda_\t{cl}l}\right),
\end{equation} 
where the clamping point is at $x=0$ (such that $u(0)=0$), the string extends to $x>0$ and $u'(+0)$ is the derivative of the solution of \eqref{eq:waveBeam} which does not satisfy the boundary condition $u'(0)=0$. The parameter $\lambda_\t{cl}$ is defined with $h$ and $\sigma$ local to the clamp\cite{bereyhi_clamp-tapering_2019}. The length of the transition region, equal to $\lambda_\t{cl}l$, is small compared to the total resonator size $l$. Nevertheless the total lossy elastic energy is commonly dominated by the part originating from the clamps\cite{unterreithmeier_damping_2010,yu_control_2012}. 

Since the lossy elastic energy (``bending energy'') is proportional to the integral of mode curvature squared, the boundary loss coefficient $\alpha$ is proportional to $(u'(+0))^2$ and the reduction of the mode gradient thereby reduces $\alpha$.

\paragraph{Propagation across a branch point ---}
An interesting situation in which suppression of the flexural mode gradient occurs is when a string mode propagates over a branch point. In order to show this, we consider a junction of three beams with rectangular cross section, highlighted by the blue contour in \figref{fig:intro}b. The dynamic equation for the two-dimensional profile of out-of-plane vibrations $u(x,y)$ is given by\cite{landau_theory_1970}
\begin{equation}\label{eq:waveMbr}
-\frac{\partial}{\partial x_i}\left(\sigma_{ij}\frac{\partial u}{\partial x_j}\right)=\omega^2\rho u,
\end{equation}  
which generalizes \eqref{eq:waveBeam}. We assume summation over the repeating indices $i$ and $j$, each of which runs over the two spatial coordinates, $x$ and $y$. The components of the stress tensor $\sigma_{ij}$ are functions of $x$ and $y$. By integrating both sides of \eqref{eq:waveMbr} over the infinitesimally small area of the blue contour in \figref{fig:intro}b and transforming the divergence into a boundary integral we find
\begin{equation}\label{eq:derivTransformIntegral}
\oint ds_i \left(\sigma_{ij}\frac{\partial u}{\partial x_j}\right) = 2\,w_{2} \sigma_{2} u_{2}'-w_1 \sigma_1 u_1'=0,
\end{equation}
where $u_1'$ and $u_2'$ are the amplitude gradients in the directions of axes $x_1$ and $x_2$, respectively. By doubling the contribution of beam two we account for the assumption that the mode branches symmetrically. Next, the balance of static tensile forces requires
\begin{equation}\label{eq:forceBal}
w_1 \sigma_1 = 2\,w_{2} \sigma_{2} \cos(\theta).
\end{equation} 
Combining \eqref{eq:derivTransformIntegral} and \eqref{eq:forceBal} we find
\begin{equation}\label{eq:uDerTrans}
u'_{2}=u'_1 \cos(\theta).
\end{equation}

\eqref{eq:uDerTrans} shows that the mode gradient is reduced by a factor of $\cos(\theta)$ after propagating over a branch point. Although the reduction in principle can be arbitrarily large if $\theta$ is close to $\pi/2$, the improvement in dissipation dilution provided by a single branch point is fairly limited. The reason is an associated increase in the distributed part of the lossy energy caused by the torsional deformation of the beams, which we will discuss in the following. We will show that cascaded branchings can greatly reduce residual lossy energy.     

\paragraph{Self-similar binary tree resonators ---}
Multiple string branchings can be cascaded such that their totality forms a binary tree, as shown in \figref{fig:intro}. After each branching the lengths of the string segments are reduced by the same ratio in order to prevent self-overlap. As realistic resonators have to be hard-clamped on all sides, we consider structures composed of two symmetric binary trees joined at the roots and clamped at the tips. We treat the case when all the strings are beams with rectangular cross section and the same thickness, a geometry that is amenable to nanofabrication. Our main qualitative results, however, are not contingent on this assumption. Since we are primarily interested in the properties of the fundamental resonator mode, in the following analysis we consider the modes that split symmetrically at each branch point. 

Binary tree resonators are convenient to analyze using a set of local axes, $x_n$, each directed along one segment, beginning at one branch point and ending at the next one as shown in \figref{fig:intro}a. Considering one path from the resonator center to one of the clamps is sufficient for describing symmetrically splitting modes. We index the branching level by $n$, and the total number of branchings is denoted by $N$. The flexural deformation of each segment as a function of the local coordinate is denoted by $u_n(x_n)$. The segment lengths, $l_n$, and widths, $w_n$ (shown in \figref{fig:intro}b), are found using the ratios $r_l$ and $r_w$ as $l_n = l_0 (r_l)^n$ and $ w_n = w_0 (r_w)^n$, respectively. Note that according to this definition the total length of the central resonator segment is $2l_0$, as it consists of two symmetric tree trunks.

Flexural modes and vibrational frequencies of tree resonators can be found by matching the solutions $u_n(x_n)$ over different segments using \eqref{eq:uDerTrans}, the continuity condition $u_n(l_n)=u_{n+1}(0)$, and the boundary conditions $u_N(l_N)=0$ and $u'_0(0)=0$ (or $u_0(0)=0$ for modes in which the two trees are deformed anti-symmetrically). With $u_n(x_n)$ in hand, one can compute the dissipation dilution factors, the $Q$s and the loss coefficients $\alpha$ and $\beta$ of the modes. However, we need to introduce one more concept before we can provide explicit expressions for these quantities. 

\begin{figure}[t]
\centering
\includegraphics[width=\columnwidth]{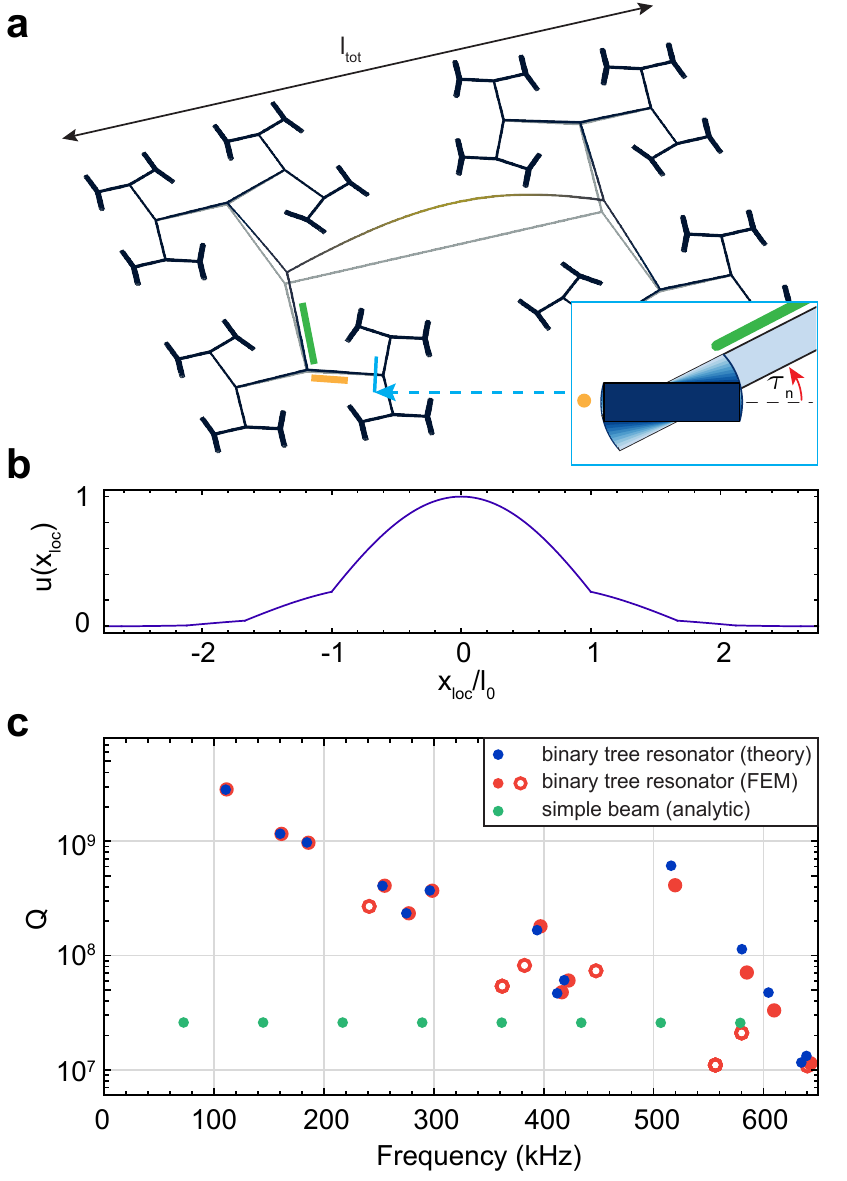}
\caption{\small{a) FEM simulation of the fundamental mode of a stress-preserving binary tree resonator with $r_l=0.67$, $l_0=1$ mm ($l_\t{tot}=3.7$ mm), $w_0=100$ nm, $h=20$ nm, $\theta =80\deg$, $N=5$. The inset schematically shows a cut view of one segment (marked with orange) and illustrates the torsion created by the previous segment (marked with green). b) The displacement of the mode shown in a) plotted over the local $x$ coordinates following a path from one tip of the tree to another. c) Quality factors and frequencies of out-of-plane modes of the resonator shown in a). Blue dots correspond to the theory presented in this work, red dots to the result of FEM simulation. Filled red dots denote symmetrically branched modes, empty dots: other modes. Green dots show out-of-plane modes of a doubly-clamped beam resonator with the same total length ($l_\t{tot}$).}}
\label{fig:sampleQs}
\end{figure}

\paragraph{Torsional lossy energy ---} 
Flexural deformations of a two-dimensional system of strings in general induce torsional deformation of the segments. If the segments have high aspect ratios, the elastic energy stored in torsion has a negligible effect on the mode frequencies, but it profoundly impacts dissipation dilution and thus quality factors. Since torsion does not produce geometrically nonlinear strain in the direction of the string axis\cite{fedorov_generalized_2019}, it only contributes to the lossy elastic energy. Below it will be shown that the torsional contribution dominates the distributed loss coefficient of binary-tree resonators in the regime of strong boundary loss suppression. 

The emergence of torsion in a tree segment is illustrated in the inset of \figref{fig:sampleQs}a. The equilibria of force moments at the junctions define the boundary conditions for the torsion angles. At the beginning of the segment, the angle is set by the previous segment as $\tau_n=u'_{n-1}(l_{n-1})\sin(\theta)$. At the end of the segment the angle is zero. The torsional energy stored by one segment is given by
\begin{equation}
\langle W_\t{tors}\rangle_n=
\frac{E w_n h^3}{6(1+\nu)}\int_{0}^{l_n}dx_n\, (\tau'(x_n))^2,
\end{equation} 
where $\nu$ is Poisson's ratio. If the aspect ratio of the segment is high (which we assume in the following), the transition from $\tau_n$ to zero happens linearly and $\tau'=\tau_n/l_n$.

Note that the torsional energy has a close relation to the ``bending'' energy of a non-uniform two-dimensional membrane. If the full resonator can be modeled as a patterned membrane, which is also true in our case, one can use the general 2D formula to find the lossy energy (see SI of Ref.~\cite{yu_control_2012}), including the contribution which we refer to as ``torsional''. Nevertheless, we find it useful to separate the torsional energy as this concept is generalizable to strings with non-rectangular cross section for which the relation to membranes is not obvious.  

\paragraph{Quality factors and loss coefficients ---} The quality factors of intrinsic loss-limited resonator modes are found by using \eqref{eq:Q}-\ref{eq:DQ}. The energies involved are calculated by summing up the contributions from all the tree segments. The lossless ``tension" energy is given by
\begin{equation}
\langle W_\t{tens}\rangle=2\sum_{n=0}^{N}2^n \sigma_n w_n h \int_{0}^{l_n}d x_n (u'_n(x_n))^2.
\end{equation}
The lossy energy consists of three contributions
\begin{equation}
\langle W_\t{lossy} \rangle=\langle W_\t{bend,b} \rangle+\langle W_\t{bend} \rangle+\langle W_\t{tors} \rangle.
\end{equation}
The distributed bending energy is
\begin{equation}
\langle W_\t{bend} \rangle=2\sum_{n=0}^{N}2^n \frac{E w_n h^3}{12} \int_{0}^{l_n}d x_n (u''_n(x_n))^2,
\end{equation}
while the boundary bending is
\begin{equation}
\langle W_\t{bend,b} \rangle= 2^{N} w_N h^2 \sqrt{\frac{E}{12}} \sqrt{\sigma_N}(u'_N(l_N-0))^2,
\end{equation}
and the torsional contribution is
\begin{equation}\label{eq:Wtors}
\langle W_\t{tors} \rangle=2\sum_{n=1}^{N}2^n \frac{E w_n h^3}{6(1+\nu)l_n} (u'_{n-1}(l_{n-1})\sin(\theta))^2.
\end{equation}
The loss coefficients in \eqref{eq:DQ} are identified as
\begin{align}
& \alpha =\frac{\langle W_\t{bend,b} \rangle}{\lambda \langle W_\t{tens}\rangle}, \\
& \beta =\frac{\langle W_\t{bend} \rangle+\langle W_\t{tors} \rangle}{\lambda^2\langle W_\t{tens}\rangle}=\beta_\t{bend}+\beta_\t{tors},
\end{align}
Note that $\alpha$ and $\beta$ are independent of $\lambda$, which in our case is defined as
\begin{equation}
\lambda=\frac{h}{l_\t{tot}}\sqrt{\frac{E}{12 \sigma_0}},
\end{equation} 
with $l_\t{tot}$ being the total resonator size in the direction along the central segment. Given $l_0$, $r_l$ and $\theta$ one can find $l_\t{tot}$ analytically, but the resulting expression is cumbersome. 

Note that unlike the case of traditional membranes and beams, the definition of $\lambda$ in our problem is a subtle question as fractal-like resonators do not have a single characteristic length scale and stress. Depending on our choice of $l$ and $\sigma$, the boundary and distributed loss coefficients $\alpha$ and $\beta$ would change, of course keeping the overall $D_Q$ constant. Our definition of $\lambda$ has the advantage that it ensures intuitive correspondence when the binary tree converges to a straight beam: if $\theta\to 0$ and $r_w\to 1/2$ then $\alpha\to 2$ and $\beta\to (m\,\pi)^2$, where $m$ is the mode order. 

\begin{figure*}[t]
\centering
\includegraphics[width=\textwidth]{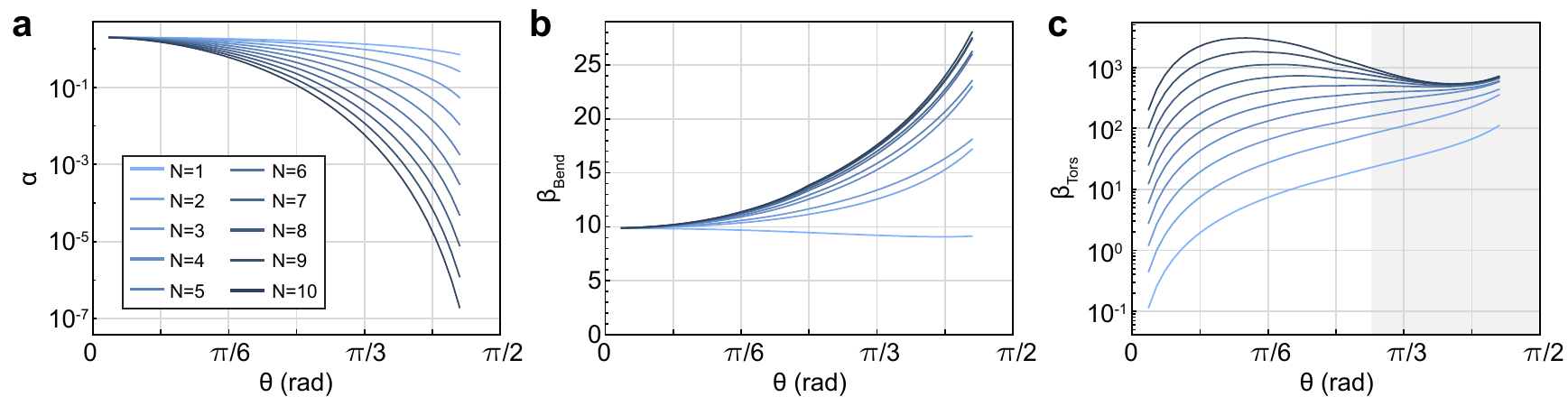}
\caption{\small{Loss coefficients for stress-preserving binary tree resonators with $r_l = r_{l.\t{crit}}(\theta)$ and different numbers of branchings $N$. a) boundary loss coefficient, b) distributed bending loss coefficient, c) distributed torsional loss coefficient, the region where $\beta_\t{tors}$ converges with increasing $N$ is shaded gray.}}
\label{fig:lossCoeffs}
\end{figure*}

\paragraph{Stress-preserving trees ---} The distribution of static stress in a binary tree resonator in general can be such that the stress is peaked either in the branch tips or in the trunk. For simplicity, we restrict our numeric analysis to the trees in which the static stress along the segments is uniform. As follows from the balance of static forces (\eqref{eq:forceBal}), the condition $\sigma_{n+1}=\sigma_{n}$ is fulfilled (and the resonator is ``stress-preserving'') if the width scaling ratio is set to $r_w= 1/(2\cos(\theta))$. If a stress-preserving resonator is patterned from a film with isotropic initial pre-stress $\sigma_\t{film}$, the static stress in all segments is given by 
\begin{equation}
\sigma_n=\sigma_\t{film}(1-\nu).
\end{equation}

\paragraph{Simulation results ---} The basic acoustic properties of binary tree resonators can be understood from an example. In \figref{fig:sampleQs} we present a simulation of the modes of a resonator made of high-stress stoichiometric silicon nitride film ($\sigma_\t{film}=1.14$ GPa, $E=250$ GPa, $\nu=0.23$, $\rho=3100$ kg/m$^3$, $1/\phi=1.4\times 10^3$) at room temperature\cite{ghadimi_strain_2017}.

The fundamental resonator mode is shown in \figref{fig:sampleQs}a and \figref{fig:sampleQs}b. The reduction of mode amplitude gradient at each branch point can be observed from these figures, together with the fact that the gradient near the clamping points approaches zero. Note, that the apparent discontinuity of mode derivative in \figref{fig:sampleQs}b is due to the turns of the path following local $x$-axes, the real two-dimensional mode has no sharp bends at the branch points.

The calculated quality factors are presented in \figref{fig:sampleQs}c, which shows that the $Q$ of the fundamental mode is enhanced by about two orders in magnitude compared to a simple doubly-clamped beam of the same size. All low-frequency flexural modes experience similar $Q$ enhancement, which gradually decreases as the acoustic wavelength becomes comparable to the length of the smallest segments. 

Two methods were used to obtain the data in \figref{fig:sampleQs}c, the theory presented in this work, which relies on the one-dimensional approximation of segment modes, and 2D finite-element method (FEM) simulation of a non-uniform plate under tension. The mode frequencies were found to agree better than within 1.5\% between the two methods in the frequency range displayed in the figure. The agreement between the quality factors is at the same level for a few lower order modes, whereas higher order modes show higher discrepancy due to the onset of hybridization between bending and torsional modes, neglected in our theoretical analysis. The FEM simulation also provides information about all the acoustic modes supported by the structure, including non-symmetrically branched and in-plane modes. For clarity we do not show in-plane modes in \figref{fig:sampleQs}c, as their quality factors are significantly lower compared to the out-of-plane modes, their density is about the same and the fundamental resonator mode never belongs to this family. 

In order to obtain a more general insight into the properties of binary tree resonators, we systematically study the variation of boundary and distributed loss coefficients of the fundamental resonator mode. These loss coefficients are material- and scale-independent and are determined by the geometric parameters $r_l$, $r_w$, $\theta$ and $N$. One of the parameters, $r_w$, is fixed to satisfy the stress-preservation condition at given $\theta$. Furthermore, we put $r_l = r_{l.\t{crit}}(\theta)$, where $r_{l.\t{crit}}$ is the value at which tip-to-tip self contact occurs in a fractal tree with infinite $N$ (and there is no self-contact for finite $N$). We sweep the remaining free parameters, $\theta$ and $N$, and present the results in \figref{fig:lossCoeffs}. It can be seen that $\alpha$ is suppressed as $\theta$ increases, while $\beta_\t{tors}$, on the contrary, goes up. Therefore, the torsional lossy energy eventually becomes the main limitation for dissipation dilution as the boundary loss is suppressed. The exact parameters at which the distributed loss matches the boundary loss, and therefore the quality factor is maximized, depend on $\lambda$. When $\lambda$ gets smaller, the optimum shifts towards larger $\theta$ or $N$ as it takes a stronger boundary energy suppression to match the distributed contribution.

\paragraph{Fractal limit ---} The data in \figref{fig:lossCoeffs} helps us understand some properties of binary tree resonators in the fractal limit, when $N$ goes to infinity. As $N$ increases, the boundary loss coefficient $\alpha$ reduces to zero and the distributed bending loss coefficient $\beta_\t{bend}$ converges to a finite value. The distributed torsion loss coefficient $\beta_\t{tors}$ has more complex behavior with increasing $N$, it can either converge to a finite value or increase indefinitely. Which of the two scenarios is realized depends on the behavior of geometric series in \eqref{eq:Wtors}, which can be shown to converge if $\cos(\theta)<\sqrt{r_l/(2r_w)}$. Correspondingly, depending on the behavior of $\beta_\t{tors}$, the $Q$ of the fundamental mode of a fractal structure can either be finite and limited by the distributed energy loss or it can be zero (i.e. the $Q$ would be low and determined by factors beyond the approximations of our theory). It is interesting to compare this conclusion to the case of membranes with self-similar boundaries, in which $Q\to 0$ was found to be the only possible scenario in the fractal limit\cite{sapoval_acoustical_1997}.

\paragraph{Conclusions and outlook ---} We showed that the boundary contribution to the lossy elastic energy of flexural modes can be suppressed in a system of tensioned strings connected to form a self-similar binary tree. This boundary loss suppression does not require the structure to extend beyond one acoustic wavelength and therefore can enhance the quality factor of the fundamental resonator mode, as well as of a multitude of other low-order modes at the same time. Our results are relevant to the design of beam\cite{verbridge_high_2006,zwickl_high_2008,unterreithmeier_universal_2009} and tethered-membrane\cite{reinhardt_ultralow-noise_2016,fischer_spin_2019} nanomechanical resonators, as well as the suspensions of macroscopic test masses\cite{corbitt_measurement_2006,corbitt_optical_2007}. The example nanoresonator design presented in this work should be amenable to nanofabrication using conventional techniques for patterning and suspending high-stress films.

Code to reproduce data in \figref{fig:sampleQs} and \figref{fig:lossCoeffs} is available on Zenodo~\cite{mathematica_package}.

\section{Acknowledgements}
The authors thank Alexander Tagantsev and Vivishek Sudhir for their comments on the manuscript. This work was supported by the Swiss National Science Foundation under grant no. 182103 and the Defense Advanced Research Projects Agency (DARPA), Defense Sciences Office (DSO), under contract no. D19AP00016 (QUORT). A.B. acknowledges support from the European Union's Horizon 2020 research and innovation program under the Marie Sklodowska-Curie grant agreement no. 722923 (OMT). N.J.E. acknowledges support from the Swiss National Science Foundation under grant no. 185870 (Ambizione). 

\bibliography{references}

\begin{thebibliography}{29}
\bibitem [{mat()}]{mathematica_package}%
  \BibitemOpen
  \href@noop {} {}\bibinfo {howpublished} {Mathematica notebook is available on
  zenodo.org, DOI:10.5281/zenodo.3496611}\BibitemShut {NoStop}%
\end{thebibliography}

\end{document}